\documentclass[twocolumn]{aastex7}

\begin{document}

\title{Enhanced Nuclear Binding Near the Proton Dripline Opens Possible Bypass of the $^{64}{\rm Ge}$ rp-process Waiting Point}

\author[orcid=0000-0002-8403-8879,gname='Zach',sname='Meisel']{Z. Meisel}
\affiliation{Department of Engineering Physics, Air Force Institute of Technology, Wright-Patterson Air Force Base, Ohio 45433, USA}
\email[show]{zachary.meisel@us.af.mil}

\author[orcid=0000-0002-0945-8654,sname='Ong']{W.-J. Ong}%
 \affiliation{%
Nuclear and Chemical Sciences Division, Lawrence Livermore National Laboratory, Livermore, California 94550, USA
}
 \email[show]{ong10@llnl.gov}

\author[orcid=0000-0001-6860-3754,gname='Jaspreet',sname='Randhawa']{J. S. Randhawa}
\affiliation{
Department of Physics and Astronomy, Mississippi State University, Mississippi State, Mississippi 39762, USA
}%
\email[show]{jsr512@msstate.edu}

\begin{abstract}

We performed astrophysics model calculations with updated nuclear data to identify a possible bypass of the $^{64}{\rm Ge}$ waiting-point, a defining feature of the rapid-proton capture (rp-) process that powers type-I x-ray bursts on accreting neutron stars. We find that the rp-process flow through the $^{64}{\rm Ge}$ bypass could be up to 36\% for astrophysically relevant conditions. Our results call for new studies of $^{65}{\rm Se}$, including the nuclear mass, $\beta$-delayed proton emission branching, and nuclear structure as it pertains to the $^{64}{\rm As}(p,\gamma)$ reaction rate at x-ray burst temperatures.

\end{abstract}

%\keywords{\uat{Galaxies}{573} --- \uat{Cosmology}{343} --- \uat{High Energy astrophysics}{739} --- \uat{Interstellar medium}{847} --- \uat{Stellar astronomy}{1583} --- \uat{Solar physics}{1476}}


\section{Introduction} 
Our understanding of atomic nuclei is often challenged by studying nuclear matter at the extremes, including the maximal asymmetry between the number of neutrons and protons in a bound nucleus that is marked by the proton drip line of the nuclear landscape~\citep{Steiner2005,Schatz2022,Watts2019}. Recent investigations of the proton drip line near nuclear mass number $A$$\approx$60--80 have revealed a number of surprising features, including slight violations of isospin mirror symmetry~\citep{Hoff2020}, accelerated two-proton decay~\citep{Wang2018}, regions of enhanced stability beyond the dripline known as sandbanks~\citep{Suzuki2017}, and hindered astrophysical nuclear burning sequences~\citep{Zhou2023}. The latter two are a direct consequence of enhanced nuclear binding relative to standard, smooth extrapolations of the nuclear mass surface, e.g. from the Atomic Mass Data collaboration's Atomic Mass Evaluation~\citep{Huang2021,Wang2021}.

Nuclear binding energies, often represented instead via the atomic mass excess ME$(Z,A)$ for a nucleus with $Z$ protons and $A$ nucleons, are a key ingredient in calculations of astrophysical burning environments~\citep{Schatz2013,Schatz2017,Meisel2020}. A nuclear reaction's energetic favorability is represented by the $Q$-value, where for radiative proton-capture $(p,\gamma)$ reactions $Q_{p,\gamma}(Z,A)={\rm ME}(Z,A)+{\rm ME}_{p}-{\rm ME}(Z+1,A+1)\equiv S_{p}(Z+1,A+1)$, where $S_{p}$ is the proton separation energy. The rapid proton capture (rp-)process that powers type-I x-ray bursts on accreting neutron star surfaces is a nuclear burning sequence consisting primarily of $(p,\gamma)$ reactions and $\beta^{+}$-decays proceeding up the nuclear landscape following $S_{p}\sim1$~MeV near the proton-drip line from roughly isotopes of calcium up to as far as isotopes of tin~\citep{Meisel2018}. The near-constant $S_{p}$ of the rp-process path is due to the high temperature, $T\sim1$~GK, and hydrogen density, $\rho_{\rm H}\sim10^{6}$~mol\,cm$^{-3}$, of the environment, which enables proton captures along an isotonic chain until $(p,\gamma)$$-$$(\gamma,p)$ equilibrium is established. At that point, nucleosynthesis stalls until the equilibrium nucleus undergoes radioactive decay. Equilibrium nuclei with half-lives $t_{1/2}$$\gtrsim$10~s are known as waiting-point nuclei and are largely responsible for the morphology of the x-ray burst light curve temporal peak and decay in luminosity~\citep{Woosley2004}.

Recent nuclear mass measurements and nuclear structure calculations established that $^{64}{\rm Ge}$ is a strong waiting-point~\citep{Lam2016,Zhou2023}, meaning that a negligible fraction of the rp-process flow proceeds through proton-capture on the abundance of $^{65}{\rm As}$ that is present due to $^{64}{\rm Ge}(p,\gamma)^{65}{\rm As}(\gamma,p)^{64}{\rm Ge}$  equilibrium ($^{64}{\rm Ge}$$\leftrightarrows$$^{65}{\rm As}$). Notably, these same measurements established that $^{66}{\rm Se}$ is more bound than previous theoretical predictions~\citep{Zhou2023}.

Here, we show that translating this assumption of increased binding to the neighboring isotope $^{65}{\rm Se}$, which is well within current estimates~\citep{Huang2021,Wang2021}, opens a possible bypass of the $^{64}{\rm Ge}$ rp-process waiting point akin to that recently established for $^{56}{\rm Ni}$~\citep{Forstner2001,Langer2014,Ong2017,Valverde2018,Saxena2022}.

\section{Bypass Physics}
\begin{figure}
    \centering
    \includegraphics[width=0.25\textwidth]{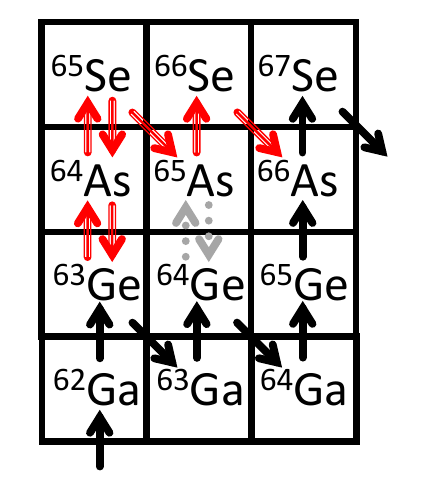}
    \caption{Primary reactions occurring in the rp-process near the $^{64}{\rm Ge}$ waiting-point, where $\uparrow$, $\downarrow$, and $\searrow$ represent $(p,\gamma)$, $(\gamma,p)$, and $\beta^{+}$-decay, respectively. Black solid arrows are the standard path, gray dotted arrows are for the (negligible) flow through the $^{64}{\rm Ge}$ waiting-point, while red striped arrows are the bypass path.}
    \label{fig:chart}
\end{figure}

The possible nuclear reaction sequences of the rp-process near $^{64}{\rm Ge}$ are shown in Fig.~\ref{fig:chart}. In the standard path, the rp-process proceeds as $^{63}{\rm Ge}(\beta^{+})^{63}{\rm Ga}(p,\gamma)^{64}{\rm Ge}(\beta^{+})^{64}{\rm Ga}(p,\gamma)^{65}{\rm Ge}(p,\gamma)^{66}{\rm As}$. Flow through the waiting-point would instead consist of $^{63}{\rm Ge}(\beta^{+})^{63}{\rm Ga}(p,\gamma)^{64}{\rm Ge}(p,\gamma)^{65}{\rm As}(p,\gamma)^{66}{\rm Se}(\beta^{+})^{66}{\rm As}$. For a bypass, the relevant reaction sequence is $^{63}{\rm Ge}(p,\gamma)^{64}{\rm As}(p,\gamma)^{65}{\rm Se}(\beta^{+})^{65}{\rm As}(p,\gamma)^{66}{\rm Se}(\beta^{+})^{66}{\rm As}$. However, the bypass can be thwarted by $^{65}{\rm Se}$ decaying via $\beta$-delayed proton ($\beta p$-)emission to $^{64}{\rm Ge}$, which is presently estimated to occur with a decay branch probability $P_{\beta p}(^{65}{\rm Se})=88^{+12}_{-13}$\%~\citep{Rogers2011}.

The portion of rp-process flow proceeding through the bypass is set by the competition between the $^{63}{\rm Ge}$ $\beta$-decay rate $\lambda_{^{63}{\rm Ge}(\beta^{+})}$ and the rate for the $^{63}{\rm Ge}(p,\gamma)^{64}{\rm As}(p,\gamma)^{65}{\rm Se}$ reaction sequence $\lambda_{2p}$. The dynamics of this reaction sequence is set by $T$, $\rho_{\rm H}$, $S_{p}(^{64}{\rm As})$, and $S_{p}(^{65}{\rm Se})$. For each of these reactions, $(p,\gamma)$$-$$(\gamma,p)$ equilibrium is established when  $(\rho_{\rm H}N_{\rm A}/n_{\gamma})(\mu_{\rm red}c^{2}/(k_{\rm B}T))^{3/2}\exp(-S_{p}/(k_{\rm B}T))\gtrsim1$, according to the Saha equation, where $\mu_{\rm red}$ is the reduced mass, $n_{\gamma}=\pi(k_{\rm B}T)^{3}/(13c^3\hbar^3)$ is the number density of photons according to the Planck distribution, and $N_{\rm A}$, $k_{\rm B}$, and $\hbar$ are the Avogadro, Boltzmann, and reduced Planck constants, respectively~\citep{Meisel2018}. Because $S_{p}(^{64}{\rm As})$$\ll$$S_{p}(^{65}{\rm Se})$, $^{63}{\rm Ge}$$\leftrightarrows$$^{64}{\rm As}$ equilibrium is established at at lower $T$ than required to establish $^{64}{\rm As}$$\leftrightarrows$$^{65}{\rm Se}$ equilibrium. 
When only $^{63}{\rm Ge}$ and $^{64}{\rm As}$ are in equilibrium, the rate from $^{63}{\rm Ge}$ into the bypass reaction sequence is  $\lambda_{2p,{\rm one}}\propto\rho_{\rm H}^{2}T^{-3/2}\exp(S_{p}(^{64}{\rm As})/(k_{\rm B}T))\langle^{64}{\rm As}(p,\gamma)\rangle$, where $\langle^{64}{\rm As}(p,\gamma)\rangle$ is the $^{64}{\rm As}(p,\gamma)$ thermonuclear reaction rate~\citep{Schatz1998}.
When both of these reactions are in equilibrium, the rate from $^{63}{\rm Ge}$ into the bypass reaction sequence is $\lambda_{2p{\rm,both}}\propto\rho_{\rm H}^{2}T^{-3}\exp((S_{p}(^{64}{\rm As})+S_{p}(^{65}{\rm Se}))/(k_{\rm B}T))\lambda_{^{65}{\rm Se}(\beta^{+})}$, where $\lambda_{^{65}{\rm Se}(\beta^{+})}$ is the $^{65}{\rm Se}$ $\beta^{+}$-decay rate~\citep{Schatz1998}. 

\begin{figure*}
    \centering
    \includegraphics[width=1\textwidth]{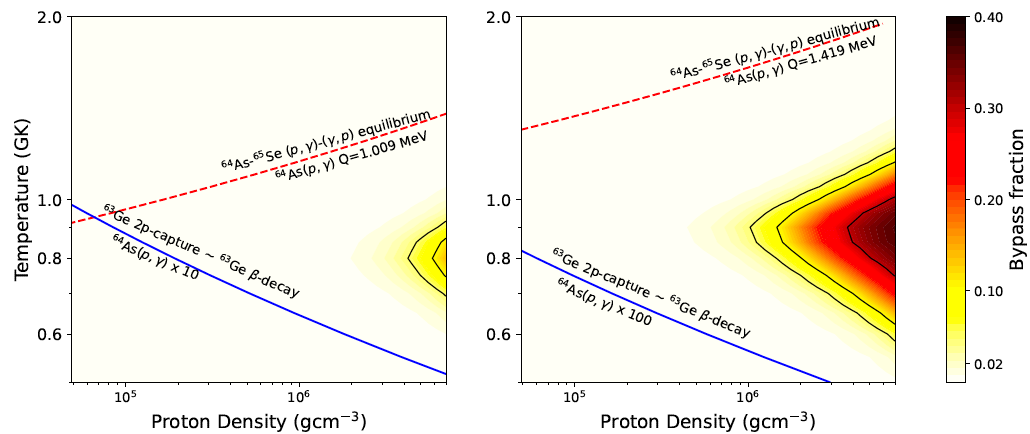}
    \caption{Fraction of rp-process flow proceeding through the $^{64}{\rm Ge}$ waiting-point bypass, as indicated by an inverted black-body radiator color intensity, assuming (left-panel) favorable
    %$\langle^{64}{\rm As}(p,\gamma)\rangle$$\times$10 and ME$(^{65}{\rm Se})$-1$\sigma$ 
    or (right-panel) maximally favorable
    %$\langle^{64}{\rm As}(p,\gamma)\rangle$$\times$100 and ME$(^{65}{\rm Se})$-2$\sigma$, with $P_{\beta p}$-2$\sigma$. 
    nuclear physics input.
    The red dashed line is the boundary for $^{64}{\rm As}$$\leftrightarrows$$^{65}{\rm Se}$ equilibrium, while the blue solid line is the boundary at which $\lambda_{2p,{\rm one}}$=$\lambda_{^{63}{\rm Ge}(\beta^{+})}$.}
    \label{fig:bypassfrac}
\end{figure*}

The boundary in $T-\rho_{\rm H}$ phase space where $\lambda_{2p,{\rm both}}$ applies is shown by the dashed-red lines of Fig.~\ref{fig:bypassfrac}. Above this boundary, no bypass is expected, as abundances along the $^{63}{\rm Ge}(p,\gamma)^{64}{\rm As}(p,\gamma)^{65}{\rm Se}$ reaction sequence are statistically distributed. Below this boundary, $\lambda_{2p,{\rm one}}$ applies. However, the flow through the bypass will only be significant if $\lambda_{2p,{\rm one}}$$\gtrsim$$\lambda_{^{63}{\rm Ge}(\beta^{+})}$, which is the boundary indicated by the solid-blue lines of Fig.~\ref{fig:bypassfrac}.  Significance of the bypass additionally requires that the amount of $^{65}{\rm As}$ created from $^{65}{\rm Se}(\beta^{+})^{65}{\rm As}$ is large relative to the $^{64}{\rm Ge}\leftrightarrows^{65}{\rm As}$ equilibrium abundance. Recent results show that this equilibrium abundance is small, but non-zero~\citep{Zhou2023}. Meanwhile, the maximum possible bypass fraction for any region of the $T-\rho_{\rm H}$ phase space is $1-P_{\beta p}(^{65}{\rm Se})$.

\section{Calculations}
In order to explore the $^{64}{\rm Ge}$ bypass physics further, we performed flow calculations following the approach previously employed for the $^{56}{\rm Ni}$ bypass~\citep{Ong2017,Saxena2022}. A limited nuclear reaction network including all $p$, $\alpha$, and $\gamma$-induced reactions as well as all $\beta^{+}$ and $\beta p$-decays on all isotopes shown in Fig.~\ref{fig:chart} was initialized with the entire non-hydrogen abundance concentrated in $^{63}{\rm Ge}$. The network was numerically solved at constant $T$ and $\rho_{\rm H}$ across the phase-space thought to be relevant for the diverse set of possible x-ray burst conditions~\citep{Parikh2008}, shown in Fig.~\ref{fig:bypassfrac}. Steady-state abundances were determined by allowing the reactions to propagate for 120~s. The ratio of the total flux of $^{66}$Se($\beta^+$) to the summed total flux of $^{66}$Se($\beta^+$), $^{65}$As($\beta^+$), and $^{64}$Ge($\beta^+$) was used to measure the fraction of rp-process flow proceeding through the $^{64}$Ge waiting-point bypass. 

Nuclear data used in the calculations includes the nuclear reaction rates of the latest REACLIB~\citep{Cyburt2010} snapshot (from 24 June 2021) updated with recently measured nuclear masses~\citep{Zhou2023} and $\beta p$-branchings~\citep{Rogers2011}, where nuclear masses were used to calculate consistent forward $(p,\gamma)$ and reverse $(\gamma,p)$ reaction rates~\citep{Rauscher2000}. Guided by the bypass physics, we investigated the impact of nuclear physics uncertainties by modifying $P_{\beta p}(^{65}{\rm Se})=88^{+12}_{-13}$\%~\citep{Rogers2011}, ME($^{65}{\rm Se})=-33020\pm300$~keV~\citep{Wang2021}, and $\langle^{64}{\rm As}(p,\gamma)\rangle$~\citep{Cyburt2010}. As the maximal bypass is of interest, we performed calculations for $P_{\beta p}$$=$62\%, i.e. a 2$\sigma$ reduction, in combination with favorable and maximally favorable values for ME($^{65}{\rm Se})$ and $\langle^{64}{\rm As}(p,\gamma)\rangle$. The favorable values correspond to ME($^{65}{\rm Se})=-33320$~keV, i.e. a 1$\sigma$ enhancement in nuclear binding, and $\langle^{64}{\rm As}(p,\gamma)\rangle$$\times$10. The maximally favorable values correspond to ME($^{65}{\rm Se})=-33620$~keV, i.e. a 2$\sigma$ enhancement in nuclear binding, and $\langle^{64}{\rm As}(p,\gamma)\rangle$$\times$100. The reaction rate variation factors are based on the evaluation of \citet{Cyburt2016}, where $\times$10 and $\times100$ were found to be the typical and extreme ratios between various theoretical predictions and experimental constraints for 
 $(p,\gamma)$ reaction rates. Note that ME($^{65}{\rm Se})$ is a theoretical estimate based on a smooth extrapolation of the local nuclear mass surface~\citep{Huang2021}.

\section{Results}
Calculated rp-process flows through the $^{64}{\rm Ge}$ waiting-point bypass are shown in Fig.~\ref{fig:bypassfrac}. The bypass begins to open for the favorable nuclear physics assumptions and achieves a maximum of 36\% for the maximally favorable nuclear physics assumptions. The contour of the $T-\rho_{\rm H}$ phase space in which the bypass is significant is set by the previously discussed equilibrium conditions. An enhancement in the nuclear binding of $^{65}{\rm Se}$ makes $^{64}{\rm As}(p,\gamma)$ relevant at lower $\rho_{\rm H}$ and higher $T$, while an enhancement of $\langle^{64}{\rm As}(p,\gamma)\rangle$ improves the competition of $^{63}{\rm Ge}(p,\gamma)^{64}{\rm As}(p,\gamma)^{65}{\rm Se}$ with $^{63}{\rm Ge}(\beta^{+})$. Even for the most favorable assumptions and astrophysical conditions, the flow cannot exceed 1-$P_{\beta p}(^{65}{\rm Se})$, as $^{65}{\rm Se}(\beta p)$ re-routes flow from the bypass into the standard rp-process path.

Relative to the $^{56}{\rm Ni}$ waiting-point bypass~\citep{Saxena2022}, the bypass around the $^{64}{\rm Ge}$ waiting-point is potentially twice as strong. However, significant flow requires plausible but somewhat more extreme $\rho_{\rm H}$. This is due  to the higher Coulomb barriers involved, as well as the much-shorter decay half-life of $^{64}{\rm Ge}$ relative to $^{56}{\rm Ni}$. The astrophysical conditions with $>$2\% bypass shown in Fig.~\ref{fig:bypassfrac} are more hydrogen-rich than thought to be relevant for the often-studied clock-burster system~\citep{Merz2021}, but are observed in calculations exploring a wider range of x-ray bursting system parameters~\citep{Schatz2017,Sultana2022}. Determining the full astrophysical implications for these and other x-ray bursting systems will require detailed studies coupling lower and higher-dimensional astrophysics model calculations with model-observation comparisons. While our results indicate that such studies are worthwhile, these involve considerable effort in their own right~\citep[e.g.][]{Cyburt2016,Meisel2018b,Meisel2022} and are left for future work.  Based on prior work investigating nuclear physics impacts for the Cu-Ni cycle~\citep{Meisel2019}, we anticipate that potential impacts could include a modified shape of the tail of the burst light curve, reducing the Gaussian fluence and non-exponentiality features, and an increased fraction of abundances above $A=64$.

\section{Discussion}
Further refining the bypass fraction for the $^{64}{\rm Ge}$ waiting-point will require dedicated studies in nuclear experiment and theory. In particular, the present work calls for a higher-precision measurement of $P_{\beta p}(^{65}{\rm Se})$, experimental determination of ME($^{65}{\rm Se})$ with precision $\mathcal{O}$(100~keV), and experimental as well as theoretical constraints on the nuclear structure of $^{65}{\rm Se}$ required to calculate the $^{64}{\rm As}(p,\gamma)$ reaction rate.

The existence of the $^{64}{\rm Ge}$ bypass relies on $P_{\beta p}(^{65}{\rm Se})<100$\%. This is not only within the present experimental uncertainty~\citep{Rogers2011}, but would be in-line with systematics. Nuclides with isospin-projection $T_{Z}=(A-2Z)/2=-3/2$ and $A=4n-1$ are expected to have similar $\beta p$-decay properties, particularly when they are within the same harmonic oscillator shell (here $fp$, i.e. $Z=20-40$)~\citep{Batchelder1993,Batchelder2020}.  Nearby nuclides in this group, $^{57}{\rm Zn}$, $^{61}{\rm Ge}$, $^{69}{\rm Kr}$, and $^{73}{\rm Sr}$, have $P_{\beta p}=$ 84.7$\pm$1.4\%, 62$\pm$4\%, 99$^{+1}_{-11}$\%, and $\sim$100\%, respectively~\citep{Saxena2022,Batchelder2020}. However, it should be noted that the $\beta^{+}$-decay children of $^{69}{\rm Kr}$ and $^{73}{\rm Sr}$ are significantly more proton unbound ($S_{p}<-600$~keV~\citep{Wang2021}) relative to $^{65}{\rm As}$ ($S_p=-220\pm40$~keV~\citep{Zhou2023}) and are thus extremely unstable to proton emission.

Enhanced nuclear binding of $^{65}{\rm Se}$ relative to a smooth nuclear mass surface is also key for the existence of the $^{64}{\rm Ge}$ bypass. Recent mass measurements of the nearby isotopes $^{64}{\rm As}$ and $^{66}{\rm Se}$ found that these are more bound than the standard smooth extrapolation of the nuclear mass surface~\citep{Huang2021,Wang2021} by 180~keV and 320~keV, respectively~\citep{Zhou2023}. These each correspond to approximately 1$\sigma$ in nuclear binding enhancement. Additionally, somewhat enhanced binding for $^{65}{\rm Se}$ is indicated when assuming isospin mirror symmetry. Empirically, $fp$-shell nuclides display a smooth trend in the Coulomb displacement energy CDE, which is a measure of isospin symmetry breaking due to the proton-neutron charge difference~\citep{Li2024}. Following the approach of \citet{Rogers2011} by applying the $^{65}{\rm As}$--$^{65}{\rm Ge}$ CDE~\citep{Li2024} to the $^{65}{\rm As}$ binding energy, correcting for the difference between empirical CDE trends for isospin 1/2 and 3/2 nuclides~\citep{Antony1997}, ME$(^{65}{\rm Se})\approx-33100$~keV. A caveat is that this estimate could be impacted by deviations from the smooth CDE trend due to odd-even staggering in $Z$~\citep{Feenberg1946}, as well as any mirror symmetry breaking~\citep[e.g.][]{Hoff2020}. 

\begin{figure}
    \centering
    \includegraphics[width=0.45\textwidth]{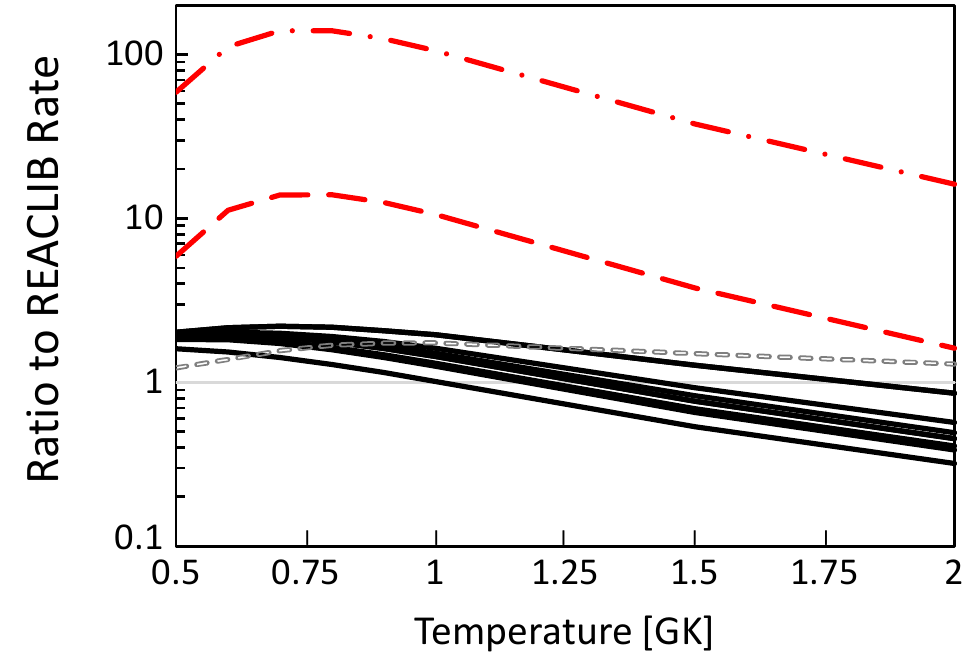}
    \caption{Ratio of $\langle^{64}{\rm As}(p,\gamma)\rangle$ from various theoretical estimates to the statistical model estimate in REACLIB~\citep{Cyburt2010}. Ratios for statistical model rates from Non-SMOKER~\citep{Rauscher2000} and TALYSv2.0 for the standard set of $\gamma$-strength functions~\citep{Koning2023} are shown by the gray open-dashed and black solid lines, respectively. Ratios for resonant rates are also shown, for a $E_{\rm r}$$=$1~MeV resonance with strength $\omega\gamma_{\rm Wig}$ (red dot-dash line) and $0.1$$\times$$\omega\gamma_{\rm Wig}$ (red long-dashed line).}
    \label{fig:rateratio}
\end{figure}

Whether the maximum bypass fraction (1-$P_{\beta p}(^{65}{\rm Se})$) is achieved for astrophysically relevant conditions depends largely upon $\langle^{64}{\rm As}(p,\gamma)\rangle$. For a $\times$100 enhancement over the REACLIB rate~\citep{Cyburt2010} and maximum $^{65}{\rm Se}$ nuclear binding, the bypass fraction can reach 36\%, while for a $\times$10 rate enhancement the bypass fraction only achieves 14\%. In the absence of experimental constraints, plausibility of these rate enhancements can only be evaluated by comparison to other theoretical predictions, as performed in Fig.~\ref{fig:rateratio}. The REACLIB rate is based on the statistical model using the Hauser-Feshbach formalism. Other commonly adopted Hauser-Feshbach estimates~\citep{Rauscher2000,Koning2023} for $\langle^{64}{\rm As}(p,\gamma)\rangle$ are within $\times$3 of the REACLIB estimate, where differences are primarily due to assumptions regarding the $\gamma$-strength function and direct capture mechanism~\citep{Wang2024}. However, a resonant rate approximation may be more valid given the relatively low nuclear level density for the excitation energies expected to be populated within $^{65}{\rm Se}$ at the relevant $T$~\citep{Rauscher1997}. For this approximation, a near-maximal rate estimate results from placing a single resonance at the optimal resonance energy $E_{\rm r}$~\citep{Newton2007} ($\approx$1~MeV in the present case) with a resonance strength near the Wigner single-particle limit $\omega\gamma_{\rm Wig}$~\citep{Teichmann1952}. In practice, even a very strong resonance strength may only be tens-of-percent of the Wigner limit, for instance as estimated for the strongest resonances of the nearby rp-process reactions $^{64}{\rm Ge}(p,\gamma)$ and $^{65}{\rm As}(p,\gamma)$~\citep{Lam2016}. Resonant reaction rates calculated~\citep{Merz2021} with these assumptions result in a $\times$$\approx$10--100 enhancement near the $T$$\approx$0.8~GK temperature at which the $^{64}{\rm Ge}$ waiting-point bypass fraction is estimated to be largest.

\section{Conclusions} 
We performed astrophysics model calculations of the reaction flow near the $^{64}{\rm Ge}$ waiting point of the rp-process that powers x-ray bursts on accreting neutron stars. Based on recently updated nuclear data near the proton dripline, we implemented somewhat enhanced nuclear binding of $^{65}{\rm Se}$ and find that this alters the textbook understanding of the rp-process  nuclear reaction sequence. For the most favorable astrophysical conditions and nuclear physics assumptions, the $^{64}{\rm Ge}$ waiting-point may be bypassed for up to 36\% of the rp-process flow. Our results call for new studies of this exotic region of isospin asymmetry in the nuclear landscape. These include nuclear mass measurements of $^{65}{\rm Se}$ with precision $\mathcal{O}$(100~keV), percent-level measurements of $P_{\beta p}(^{65}{\rm Se})$, and studies of $^{65}{\rm Se}$ nuclear structure as it pertains to the $^{64}{\rm As}(p,\gamma)$ reaction rate at $T$$\approx$0.8~GK.

%% Please use the acknowledgment and contribution environments. This will 
%% be anonomyized when the "anonymous" style option is used. 
\begin{acknowledgments}
The views expressed in this article are those of the authors and do not reflect the official guidance or position of the United States Government, the Department of Defense, the United States Air Force, or the United States Space Force. This work was performed under the auspices of the U.S. Department of Energy by Lawrence Livermore National Laboratory under Contract No. DE-AC52-07NA27344. JSR acknowledges the support from Office of Research and Economic Development, and College of Arts and Sciences at Mississippi State University.
\end{acknowledgments}

%% For this sample we use BibTeX plus aasjournalv7.bst to generate the

\bibliography{mainarticle_apj}{}

\providecommand{\noopsort}[1]{}\providecommand{\singleletter}[1]{#1}%
\begin{thebibliography}{}
\expandafter\ifx\csname natexlab\endcsname\relax\def\natexlab#1{#1}\fi
\providecommand{\url}[1]{\href{#1}{#1}}
\providecommand{\dodoi}[1]{doi:~\href{http://doi.org/#1}{\nolinkurl{#1}}}
\providecommand{\doeprint}[1]{\href{http://ascl.net/#1}{\nolinkurl{http://ascl.net/#1}}}
\providecommand{\doarXiv}[1]{\href{https://arxiv.org/abs/#1}{\nolinkurl{https://arxiv.org/abs/#1}}}

\bibitem[{M.~S. {Antony} {et~al.}(1997){Antony}, {Pape}, \& {Britz}}]{Antony1997}
{Antony}, M.~S., {Pape}, A., \& {Britz}, J. 1997, \bibinfo{title}{{Coulomb Displacement Energies between Analog Levels for 3$\leq$$A$$\leq$239},} Atom. Dat. Nucl. Dat. Tab., 66, 1, \dodoi{10.1006/adnd.1997.0740}

\bibitem[{J.~C. {Batchelder}(2020){Batchelder}}]{Batchelder2020}
{Batchelder}, J.~C. 2020, \bibinfo{title}{{Recommended values for {\ensuremath{\beta}}$^{+}$-delayed proton and {\ensuremath{\alpha}} emission},} Atom. Dat. Nucl. Dat. Tab., 132, 101323, \dodoi{10.1016/j.adt.2019.101323}

\bibitem[{J.~C. {Batchelder} {et~al.}(1993){Batchelder}, {Moltz}, {Ognibene}, {Rowe}, \& {Cerny}}]{Batchelder1993}
{Batchelder}, J.~C., {Moltz}, D.~M., {Ognibene}, T.~J., {Rowe}, M.~W., \& {Cerny}, J. 1993, \bibinfo{title}{{Beta-delayed proton decay of $^{65}$Se},} Phys. Rev. C, 47, 2038, \dodoi{10.1103/PhysRevC.47.2038}

\bibitem[{R.~H. {Cyburt} {et~al.}(2016){Cyburt}, {Amthor}, {Heger}, {Johnson}, {Keek}, {Meisel}, {Schatz}, \& {Smith}}]{Cyburt2016}
{Cyburt}, R.~H., {Amthor}, A.~M., {Heger}, A., {et~al.} 2016, \bibinfo{title}{{Dependence of X-Ray Burst Models on Nuclear Reaction Rates},} \apj, 830, 55, \dodoi{10.3847/0004-637X/830/2/55}

\bibitem[{R.~H. {Cyburt} {et~al.}(2010){Cyburt} {et~al.}}]{Cyburt2010}
{Cyburt}, R.~H., {et~al.} 2010, \bibinfo{title}{{The JINA REACLIB Database: Its Recent Updates and Impact on Type-I X-ray Bursts},} \apjs, 189, 240, \dodoi{10.1088/0067-0049/189/1/240}

\bibitem[{E. {Feenberg} \& G. {Goertzel}(1946){Feenberg} \& {Goertzel}}]{Feenberg1946}
{Feenberg}, E., \& {Goertzel}, G. 1946, \bibinfo{title}{{Theory of Nuclear Coulomb Energy},} Phys. Rev., 70, 597, \dodoi{10.1103/PhysRev.70.597}

\bibitem[{O. {Forstner} {et~al.}(2001){Forstner}, {Herndl}, {Oberhummer}, {Schatz}, \& {Brown}}]{Forstner2001}
{Forstner}, O., {Herndl}, H., {Oberhummer}, H., {Schatz}, H., \& {Brown}, B.~A. 2001, \bibinfo{title}{{Thermonuclear reaction rate of $^{56}$Ni(p,{\ensuremath{\gamma}})$^{57}$Cu and $^{57}$Cu(p,{\ensuremath{\gamma}})$^{58}$Zn},} \prc, 64, 045801, \dodoi{10.1103/PhysRevC.64.045801}

\bibitem[{D.~E.~M. {Hoff} {et~al.}(2020){Hoff} {et~al.}}]{Hoff2020}
{Hoff}, D.~E.~M., {et~al.} 2020, \bibinfo{title}{{Mirror-symmetry violation in bound nuclear ground states},} \nat, 580, 52, \dodoi{10.1038/s41586-020-2123-1}

\bibitem[{W. Huang {et~al.}(2021)Huang, Wang, Kondev, Audi, \& Naimi}]{Huang2021}
Huang, W., Wang, M., Kondev, F., Audi, G., \& Naimi, S. 2021, \bibinfo{title}{The {AME} 2020 atomic mass evaluation (I). Evaluation of input data, and adjustment procedures,} Chinese Phys. C, 45, 030002, \dodoi{10.1088/1674-1137/abddb0}

\bibitem[{A. {Koning} {et~al.}(2023){Koning}, {Hilaire}, \& {Goriely}}]{Koning2023}
{Koning}, A., {Hilaire}, S., \& {Goriely}, S. 2023, \bibinfo{title}{{TALYS: modeling of nuclear reactions},} Eur. Phys. J. A, 59, 131, \dodoi{10.1140/epja/s10050-023-01034-3}

\bibitem[{Y.~H. {Lam} {et~al.}(2016){Lam} {et~al.}}]{Lam2016}
{Lam}, Y.~H., {et~al.} 2016, \bibinfo{title}{{Reaction Rates of $^{64}$Ge$(p,\gamma)^{65}$As and $^{65}$As$(p,\gamma)^{66}$Se and the Extent of Nucleosynthesis in Type I X-Ray Bursts},} \apj, 818, 78, \dodoi{10.3847/0004-637X/818/1/78}

\bibitem[{C. {Langer} {et~al.}(2014){Langer} {et~al.}}]{Langer2014}
{Langer}, C., {et~al.} 2014, \bibinfo{title}{{Determining the rp-Process Flow through $^{56}{\rm Ni}$: Resonances in $^{57}{\rm Cu}(p,\gamma)^{58}{\rm Ni}$ Identified with GRETINA},} \prl, 113, 032502, \dodoi{10.1103/PhysRevLett.113.032502}

\bibitem[{H.~F. Li {et~al.}(2024)Li {et~al.}}]{Li2024}
Li, H.~F., {et~al.} 2024, \bibinfo{title}{{Exploring isospin-nonconserving effects in the upper fp shell with new mass measurements},} Phys. Rev. C Lett., 110, L021301, \dodoi{10.1103/PhysRevC.110.L021301}

\bibitem[{Z. {Meisel}(2018){Meisel}}]{Meisel2018b}
{Meisel}, Z. 2018, \bibinfo{title}{{Consistent Modeling of GS 1826-24 X-Ray Bursts for Multiple Accretion Rates Demonstrates the Possibility of Constraining rp-process Reaction Rates},} \apj, 860, 147, \dodoi{10.3847/1538-4357/aac3d3}

\bibitem[{Z. {Meisel}(2020){Meisel}}]{Meisel2020}
{Meisel}, Z. 2020, \bibinfo{title}{{Mapping the frontiers of the nuclear mass surface},} J. Phys.: Conf. Ser., 1668, 012026, \dodoi{10.1088/1742-6596/1668/1/012026}

\bibitem[{Z. Meisel {et~al.}(2018)Meisel, Deibel, Keek, Shternin, \& Elfritz}]{Meisel2018}
Meisel, Z., Deibel, A., Keek, L., Shternin, P., \& Elfritz, J. 2018, \bibinfo{title}{Nuclear Physics of the Outer Layers of Accreting Neutron Stars,} J. Phys. G, 45, 093001, \dodoi{10.1088/1361-6471/aad171}

\bibitem[{Z. {Meisel} {et~al.}(2019){Meisel}, {Merz}, \& {Medvid}}]{Meisel2019}
{Meisel}, Z., {Merz}, G., \& {Medvid}, S. 2019, \bibinfo{title}{{Influence of Nuclear Reaction Rate Uncertainties on Neutron Star Properties Extracted from X-Ray Burst Model-Observation Comparisons},} \apj, 872, 84, \dodoi{10.3847/1538-4357/aafede}

\bibitem[{Z. {Meisel} {et~al.}(2022){Meisel} {et~al.}}]{Meisel2022}
{Meisel}, Z., {et~al.} 2022, \bibinfo{title}{{Improved nuclear physics near A =61 refines urca neutrino luminosities in accreted neutron star crusts},} \prc, 105, 025804, \dodoi{10.1103/PhysRevC.105.025804}

\bibitem[{G. {Merz} \& Z. {Meisel}(2021){Merz} \& {Meisel}}]{Merz2021}
{Merz}, G., \& {Meisel}, Z. 2021, \bibinfo{title}{{Urca nuclide production in Type-I X-ray bursts and implications for nuclear physics studies},} Mon. Not. R. Astron. Soc., 500, 2958, \dodoi{10.1093/mnras/staa3414}

\bibitem[{J.~R. {Newton} {et~al.}(2007){Newton}, {Iliadis}, {Champagne}, {Coc}, {Parpottas}, \& {Ugalde}}]{Newton2007}
{Newton}, J.~R., {Iliadis}, C., {Champagne}, A.~E., {et~al.} 2007, \bibinfo{title}{{Gamow peak in thermonuclear reactions at high temperatures},} Phys. Rev. C, 75, 045801, \dodoi{10.1103/PhysRevC.75.045801}

\bibitem[{W.~J. {Ong} {et~al.}(2017){Ong} {et~al.}}]{Ong2017}
{Ong}, W.~J., {et~al.} 2017, \bibinfo{title}{{Low-lying level structure of $^{56}$Cu and its implications for the rp process},} \prc, 95, 055806, \dodoi{10.1103/PhysRevC.95.055806}

\bibitem[{A. {Parikh} {et~al.}(2008){Parikh}, {Jos{\'e}}, {Moreno}, \& {Iliadis}}]{Parikh2008}
{Parikh}, A., {Jos{\'e}}, J., {Moreno}, F., \& {Iliadis}, C. 2008, \bibinfo{title}{{The Effects of Variations in Nuclear Processes on Type I X-Ray Burst Nucleosynthesis},} \apjs, 178, 110, \dodoi{10.1086/589879}

\bibitem[{T. {Rauscher} \& F.-K. {Thielemann}(2000){Rauscher} \& {Thielemann}}]{Rauscher2000}
{Rauscher}, T., \& {Thielemann}, F.-K. 2000, \bibinfo{title}{{Astrophysical Reaction Rates From Statistical Model Calculations},} Atom. Dat. Nucl. Dat. Tab., 75, 1, \dodoi{10.1006/adnd.2000.0834}

\bibitem[{T. {Rauscher} {et~al.}(1997){Rauscher}, {Thielemann}, \& {Kratz}}]{Rauscher1997}
{Rauscher}, T., {Thielemann}, F.-K., \& {Kratz}, K.-L. 1997, \bibinfo{title}{{Nuclear level density and the determination of thermonuclear rates for astrophysics},} Phys. Rev. C, 56, 1613, \dodoi{10.1103/PhysRevC.56.1613}

\bibitem[{A.~M. {Rogers} {et~al.}(2011){Rogers} {et~al.}}]{Rogers2011}
{Rogers}, A.~M., {et~al.} 2011, \bibinfo{title}{{$^{69}$Kr {\ensuremath{\beta}}-delayed proton emission: A Trojan horse for studying states in proton-unbound $^{69}$Br},} Phys. Rev. C Rapid Comm., 84, 051306, \dodoi{10.1103/PhysRevC.84.051306}

\bibitem[{M. {Saxena} {et~al.}(2022){Saxena} {et~al.}}]{Saxena2022}
{Saxena}, M., {et~al.} 2022, \bibinfo{title}{{$^{57}$Zn {\ensuremath{\beta}}-delayed proton emission establishes the $^{56}$Ni rp-process waiting point bypass},} Phys. Lett. B, 829, 137059, \dodoi{10.1016/j.physletb.2022.137059}

\bibitem[{H. {Schatz}(2013){Schatz}}]{Schatz2013}
{Schatz}, H. 2013, \bibinfo{title}{{Nuclear masses in astrophysics},} Int. J. Mass Spectrom., 349-350, 181, \dodoi{10.1016/j.ijms.2013.03.016}

\bibitem[{H. {Schatz} \& W.~J. {Ong}(2017){Schatz} \& {Ong}}]{Schatz2017}
{Schatz}, H., \& {Ong}, W.~J. 2017, \bibinfo{title}{{Dependence of X-Ray Burst Models on Nuclear Masses},} Astrophys. J., 844, 139, \dodoi{10.3847/1538-4357/aa7de9}

\bibitem[{H. {Schatz} {et~al.}(1998){Schatz} {et~al.}}]{Schatz1998}
{Schatz}, H., {et~al.} 1998, \bibinfo{title}{{rp-Process Nucleosynthesis at Extreme Temperature and Density Conditions},} Phys. Rep., 294, 167, \dodoi{10.1016/S0370-1573(97)00048-3}

\bibitem[{H. {Schatz} {et~al.}(2022){Schatz} {et~al.}}]{Schatz2022}
{Schatz}, H., {et~al.} 2022, \bibinfo{title}{{Horizons: nuclear astrophysics in the 2020s and beyond},} J. Phys. G, 49, 110502, \dodoi{10.1088/1361-6471/ac8890}

\bibitem[{A.~W. {Steiner} {et~al.}(2005){Steiner}, {Prakash}, {Lattimer}, \& {Ellis}}]{Steiner2005}
{Steiner}, A.~W., {Prakash}, M., {Lattimer}, J.~M., \& {Ellis}, P.~J. 2005, \bibinfo{title}{{Isospin asymmetry in nuclei and neutron stars},} Phys. Rep., 411, 325, \dodoi{10.1016/j.physrep.2005.02.004}

\bibitem[{I. {Sultana} {et~al.}(2022){Sultana}, {Estrad{\'e}}, {Borowiak}, {Elliott}, {Meyer}, \& {Schatz}}]{Sultana2022}
{Sultana}, I., {Estrad{\'e}}, A., {Borowiak}, J., {et~al.} 2022, \bibinfo{title}{{Sensitivity Study of Type-I X-ray Bursts to Nuclear Reaction Rates},} Euro. Phys. J. Web Conf., 260, 11040, \dodoi{10.1051/epjconf/202226011040}

\bibitem[{H. {Suzuki} {et~al.}(2017){Suzuki} {et~al.}}]{Suzuki2017}
{Suzuki}, H., {et~al.} 2017, \bibinfo{title}{{Discovery of $^{72}$Rb: A Nuclear Sandbank Beyond the Proton Drip Line},} \prl, 119, 192503, \dodoi{10.1103/PhysRevLett.119.192503}

\bibitem[{T. {Teichmann} \& E.~P. {Wigner}(1952){Teichmann} \& {Wigner}}]{Teichmann1952}
{Teichmann}, T., \& {Wigner}, E.~P. 1952, \bibinfo{title}{{Sum Rules in the Dispersion Theory of Nuclear Reactions},} Phys. Rev., 87, 123, \dodoi{10.1103/PhysRev.87.123}

\bibitem[{A.~A. {Valverde} {et~al.}(2018){Valverde} {et~al.}}]{Valverde2018}
{Valverde}, A.~A., {et~al.} 2018, \bibinfo{title}{{High-Precision Mass Measurement of $^{56}{\rm Cu}$ and the Redirection of the rp-Process Flow},} \prl, 120, 032701, \dodoi{10.1103/PhysRevLett.120.032701}

\bibitem[{B. {Wang} {et~al.}(2024){Wang}, {Xu}, \& {Goriely}}]{Wang2024}
{Wang}, B., {Xu}, Y., \& {Goriely}, S. 2024, \bibinfo{title}{{Systematic study of the radiative proton capture including the compound, pre-equilibrium, and direct mechanisms},} Phys. Rev. C, 109, 014611, \dodoi{10.1103/PhysRevC.109.014611}

\bibitem[{M. Wang {et~al.}(2021)Wang, Huang, Kondev, Audi, \& Naimi}]{Wang2021}
Wang, M., Huang, W., Kondev, F., Audi, G., \& Naimi, S. 2021, \bibinfo{title}{The {AME} 2020 atomic mass evaluation ({II}). Tables, graphs and references,} Chinese Phys. C, 45, 030003, \dodoi{10.1088/1674-1137/abddaf}

\bibitem[{S.~M. {Wang} \& W. {Nazarewicz}(2018){Wang} \& {Nazarewicz}}]{Wang2018}
{Wang}, S.~M., \& {Nazarewicz}, W. 2018, \bibinfo{title}{{Puzzling Two-Proton Decay of $^{67}$Kr},} \prl, 120, 212502, \dodoi{10.1103/PhysRevLett.120.212502}

\bibitem[{A.~L. {Watts} {et~al.}(2019){Watts} {et~al.}}]{Watts2019}
{Watts}, A.~L., {et~al.} 2019, \bibinfo{title}{{Dense matter with eXTP},} Science China Physics, Mechanics, and Astronomy, 62, 29503, \dodoi{10.1007/s11433-017-9188-4}

\bibitem[{S.~E. {Woosley} {et~al.}(2004){Woosley}, {Heger}, {Cumming}, {Hoffman}, {Pruet}, {Rauscher}, {Fisker}, {Schatz}, {Brown}, \& {Wiescher}}]{Woosley2004}
{Woosley}, S.~E., {Heger}, A., {Cumming}, A., {et~al.} 2004, \bibinfo{title}{{Models for Type I X-Ray Bursts with Improved Nuclear Physics},} \apjs, 151, 75, \dodoi{10.1086/381533}

\bibitem[{X. {Zhou} {et~al.}(2023){Zhou} {et~al.}}]{Zhou2023}
{Zhou}, X., {et~al.} 2023, \bibinfo{title}{{Mass measurements show slowdown of rapid proton capture process at waiting-point nucleus $^{64}$Ge},} Nature Physics, 19, 1091, \dodoi{10.1038/s41567-023-02034-2}

\end{thebibliography}
\bibliographystyle{aasjournalv7}



\end{document}